\documentstyle[aps,epsf,prc]{revtex}

\begin{document}
\title{Threshold open charm production in nucleon-nucleon collisions}
\author{Michail P. Rekalo \footnote{ Permanent address:
\it National Science Center KFTI, 310108 Kharkov, Ukraine}
}
\address{Middle East Technical University, 
Physics Department, Ankara 06531, Turkey}
\author{Egle Tomasi-Gustafsson}
\address{\it DAPNIA/SPhN, CEA/Saclay, 91191 Gif-sur-Yvette Cedex, 
France}
\date{\today}

\maketitle
\begin{abstract}
The associative charm particles production in nucleon-nucleon collisions  $N+N\to \Lambda_c(\Sigma_c)+\overline{D}+N$, is described in a general way and  the spin and isospin structure of the corresponding matrix elements are derived. Using an analogy with strange particle ($\Lambda K$)-production, the $D-$meson exchange model is considered in detail. Estimations of the energy behavior of the threshold cross sections show a large dependence on the form of the phenomenological hadron form factors and indicate that, at threshold, the cross section is three order of magnitude smaller than for strange particle production.
\end{abstract}

\section{Introduction}

New  experimental facilities, planned in the near future \cite{GSI}, will be able to make detailed measurement of open charm production  in $pN$-, $NA$- or $AA$-collisions, in a wide energy region, starting from threshold. One can expect that the physics of close-to-threshold $\Lambda_cD$-production is similar to $\Lambda K$-production.

Last precise experiments about strange particle production in $pp$-collisions gave interesting information about the possible reaction mechanism \cite{Mo02}. For example, the relative value of the total cross sections for  $\Lambda K$ and $\Sigma^0 K$-production, and in particular the large and negative polarization transfer coefficient $D_{nn}$ (from the polarized proton beam to the produced $\Lambda$-hyperon) can be considered a strong indication of $K$-meson exchange as the main reaction mechanism - in this energy region. The final $\Lambda N$-interaction strongly affects the different polarization observables, suggesting a sensitive method for the determination of low-energy parameters of the $\Lambda N$ or $\Sigma N$-interaction. Such information is useful for the determination of the $\Lambda N$ and $\Sigma N$-potential and therefore for predictions of the properties of hypernuclei \cite{Hyp00}.

The existing data about $N+N\to \Lambda(\Sigma)+K+N$, in the near threshold region, do not exclude a contribution of $\pi$-exchange, in particular in the case of $\Sigma$-production, where the $K$-exchange is depressed by a smaller $g_{p\Sigma K}$-coupling constant (following $SU_3$-symmetry one finds $g^2_{\Lambda p K}/g^2_{\Sigma p K} \simeq 27$).

The experience gained in the theoretical and experimental study of strange particles can be extrapolated to the case of the processes $N+N\to \Lambda_c(\Sigma_c)+\overline{D}+N$, in the threshold region. No experimental data exist here: the lowest energy, where open charm production has been detected in $pp-$collisions is $E_p$=70 GeV ($\sqrt{s}$=11 GeV) \cite{Am01}. The experimental activity is focussed on QCD-inspired approaches, mainly for inclusive charm production, which can not be easily extrapolated to the threshold region. In a more traditional way, the process $p+p\to \Lambda_c+\overline{D}+N$ has been considered twenty years ago, in the framework of one-pion exchange model \cite{Bo83}, using a specific approach, for the calculation of the amplitudes for such elementary subprocess, as $\pi+N\to \Lambda_c+D$ or $\pi+N\to \Lambda_c+D^*$. This approach is based on the Quark-Gluon-String model, which can be considered as the non-perturbative topological equivalent of the corresponding Regge phenomenology. Note that, in such approach, the spin structure of the amplitudes for $\pi +N\to \Lambda_c+D(D^*)$-process is ignored, and in principle, it is crucial for a correct calculation in the near threshold region. Moreover, a number of parameters is necessary for such calculations, concerning, in particular, the $D^*$-Regge trajectory.

The understanding of elementary open charm production processes in $\pi N$ or $NN$-collisions is a very actual problem,  in connection with leptonic pair production in $Pb+Pb$-collisions \cite{Ab00} at SPS. The intensive charm production in secondary $\pi N$-collisions \cite{Ca01a} can be considered an important source of leptonic pairs in the intermediate mass region \cite{Bo99}.

For a reliable description of the overall charm dynamics of $AA$-collisions, it is necessary to have a good parametrization of the  elementary processes of open charm production in $\pi N$ and $NN$-collisions. The parametrization of the energy behavior of the total cross section for different processes of $D$ and $\overline{D}$-production  for $\sqrt{s}\ge$ 10 GeV generally used in simulation codes, can not be applied near threshold as it violates the general threshold behavior for two- or three-particle production. We completely agree on the  explicit statement of  Ref. \cite{Ca01b}:'{\it 
We have to point out that our parametrizations for the
differential and total cross sections for open charm (as well as
charmonia) become questionable at low energy, but also at high
energy. It is thus mandatory that they have to be controlled by
experimental data from $pp$, $pA$ and $\pi N$ reactions before
reliable conclusions on open charm dynamics in nucleus-nucleus
reactions can be drawn.}' 

In this paper we analyze the $N+N\to \Lambda_c(\Sigma_c)+\overline{D}+N$-processes, in the near threshold region, taking rigorously into account the spin and isospin structure of the corresponding matrix elements. In our analysis, we follow a very close analogy between open charm and open strangeness production in nucleon-nucleon collisions. According to the  $SU_4$-symmetry (which gives useful relations for meson-baryon coupling constants, such as $g_{p\Lambda K^+}=g_{p \Lambda_c^+ D^0}$ and $g_{p \Sigma^0  K^+} = g_{p \Sigma_c^0  \overline{D^0}}$), one can expect that namely the $D$-exchange is important for the processes 
$N+N\to \Lambda_c(\Sigma_c)+\overline{D}+N$ in the near threshold region. The amplitude for such mechanism can be predicted, being driven by elastic $\overline{D}N$-scattering (at zero energy), in analogy to elastic $KN$-scattering, with real scattering lengths. The possible $\pi$-exchange - involving the elementary subprocess $\pi +N\to \Lambda_c+D$, seems more complicated and difficult to estimate (in model independent way).

This paper is organized as follows. In Section II we establish the spin and isospin structure of the threshold matrix elements  for the processes 
$N+N\to \Lambda_c(\Sigma_c)+\overline{D}+N$. The calculation of the threshold partial amplitudes for $\Lambda_c\overline{D}$-production in the framework of 
$\overline{D}$-exchange is done in Section III. The estimation of the energy dependence of the total cross section is done in Section IV.

\section{Spin and isospin in the processes $N+N\to \Lambda_c(\Sigma_c)+\overline{D}+N$}
The strong correlation between the spin and isospin structures of the threshold matrix element for the $N+N\to \Lambda_c(\Sigma_c)+\overline{D}+N$ reactions is an important characteristic of the threshold $NN$-dynamics. First of all, this correlation  causes the absence of factorization of spin and isospin variables, which is typical for the majority of hadronic processes (see $\pi N$-scattering, for example). Factorization means that we can consider the spin structure of the matrix elements independently from its isospin structure. Moreover, all possible spin structures, which are allowed by the symmetry properties of the strong interaction for the considered matrix element, contain, in general, all possible isospin structures. It is not the case for the threshold matrix element of the  $N+N\to \Lambda_c(\Sigma_c)+\overline{D}+N$-processes, where there is a strong and not trivial connection of the spin and isospin variables. As a result, different spin amplitudes have different isotopic structures.

Using the isotopic invariance of the strong interaction and taking into account that $I(\Sigma_c)=1$, $I(\overline{D})=1/2$, one can find for $N+N\to\Sigma_c+\overline{D}+N$:
\begin{equation}
\begin{array}{rrrr}
{\cal M}(pp\to \Sigma_c^{++}D^- p)=
&A_{11}&+\sqrt{2}A_{10}, &  \\
{\cal M}(pp\to \Sigma_c^{++}\overline{D^0} n)=
&A_{11}&-\sqrt{2}A_{10},&  \nonumber \\
{\cal M}(pp\to \Sigma_c^+\overline{D^0} p)=
&-\sqrt{2}A_{11}, & & \nonumber\\
{\cal M}(np\to \Sigma_c^{++}D^- n )=
&~A_{11}& &+\sqrt{2}A_{01},\label{eq:amp} \\
{\cal M}(np\to \Sigma_c^+\overline{D^0} p)=
&-A_{11}&& +\sqrt{2}A_{01},\nonumber\\
{\cal M}(np\to \Sigma_c^+ D^- p)=&&A_{10} &-A_{01}, \nonumber\\
{\cal M}(np\to \Sigma_c^+\overline{D^0} n)=&&-A_{10} &-A_{01}, \nonumber
\end{array}
\end{equation} 
where $A_{I_1 I_2}$ are the isotopic amplitudes, corresponding to the total isospin $I_1$ for the initial nucleons and the total isospin $I_2$ for the produced $\overline{D}N$-system.

One can see, from Eq. (\ref{eq:amp}), that seven different matrix elements for 
$N+N\to\Sigma_c+\overline{D}+N$, are characterized by three complex isotopic amplitudes: $A_{11}$, $A_{10}$ and $A_{01}$. Therefore, in the general case, taking into account the possible interference between these amplitudes, one can find the following relation between the differential cross sections:
$$
\displaystyle\frac{d\sigma}{d\omega}(pp\to \Sigma_c^{++}D^- p)+\displaystyle\frac{d\sigma}{d\omega}(pp\to \Sigma_c^{++}\overline{D^0} n)+\displaystyle\frac{d\sigma}{d\omega}(np\to \Sigma_c^{++}D^- n )+\displaystyle\frac{d\sigma}{d\omega}(np\to \Sigma_c^+\overline{D^0} p)=$$
\begin{equation}
=2\left [\displaystyle\frac{d\sigma}{d\omega}(pp\to \Sigma_c^+\overline{D^0} p)+
\displaystyle\frac{d\sigma}{d\omega}(np\to \Sigma_c^+ D^- p)+
\displaystyle\frac{d\sigma}{d\omega}(np\to \Sigma_c^+\overline{D^0} n)\right ], \label{eq:sig}
\end{equation} 
where $d{\omega}$ is the phase space element for three-particle final state. Note that Eq. (\ref{eq:sig}) holds in any kinematical condition and for any reaction mechanism.

The situation essentially changes near the reaction threshold, where all final particles are produced in $s$-state. The selection rules with respect to the Pauli principle, the P-invariance and the conservation of isospin and of the total angular momentum, allow to parametrize the spin structure of the matrix element for $pp$ and $np$-collisions as \cite{Re97}:
\begin{equation}
{\cal M}(pp)=f_{10}(\tilde{\chi}_2~\sigma_y ~\vec{\sigma}
\cdot\hat{\vec k }\chi_1)~(\chi^{\dagger}_4 \sigma_y\ \tilde{\chi}^{\dagger}_3 )+i f_{11}[\tilde{\chi}_2~\sigma_y(\vec\sigma\times\hat{\vec k})_a\chi_1]~(\chi^{\dagger}_4 \sigma_a\sigma_y\tilde{\chi}^{\dagger}_3 ),
\label{eq:mpp} 
\end{equation}
\begin{eqnarray}
&{\cal M}(np)=&f_{01}(\tilde{\chi}_2~\sigma_y \chi_1)~(\chi^{\dagger}_4 
\vec{\sigma}\cdot\hat{\vec k }\sigma_y\tilde{\chi}^{\dagger}_3 )+f_{10}(\tilde{\chi}_2\sigma_y\vec{\sigma}\cdot\hat{\vec k }\chi_1)(\chi^{\dagger}_4
\sigma_y \tilde{\chi}^{\dagger}_3 ) \nonumber \\
&&+if_{11}[\tilde{\chi}_2~\sigma_y (\vec\sigma\times\hat{\vec k})_a\chi_1]~(\chi^{\dagger}_4 \sigma_a\sigma_y\tilde{\chi}^{\dagger}_3 ),
\label{eq:mnp} 
\end{eqnarray}
where we are using the following notations:
$\hat{\vec k}$ is the unit vector along the 3-momentum of the initial nucleon; $\chi_1$ and $\chi_2$ are the
two-component spinors of the colliding nucleons, $\chi_3$ and $\chi_4$ are the
two-component spinors of the scattered nucleon and of the produced charmed baryon; $\sigma_a=(\sigma_x,\sigma_y,\sigma_z)$ are the standard 2$\times$ 2 Pauli matrices, $f_{01}$ is the singlet-triplet amplitude for ${\cal B}_c=\Sigma_c$ or $\Lambda_c$ threshold production, $f_{11}(f_{10})$ is the triplet-triplet (triplet-singlet) amplitude. One can see that the amplitude $f_{01}$ is proportional to the isotopic amplitude $A_{01}$. Moreover, the isotopic amplitudes $A_{11}$ and $A_{10}$ have the same spin structure as in Eq. (\ref{eq:mpp}).  A direct consequence of the threshold spin structure, Eqs. (\ref{eq:mpp}) and (\ref{eq:mnp}), is that the interferences $A_{11}\bigotimes A_{01}^*$ and $A_{10}\bigotimes A_{01}^*$, which can be present in the general case, vanish in the threshold region, after summing over the polarizations of the colliding nucleons. Therefore we can derive the following  relations, additional to (\ref{eq:sig}), between the differential cross sections, which hold for any model, at threshold:
$$ \displaystyle\frac{d\sigma}{d\omega}(np\to \Sigma_c^{++}D^- n )= \displaystyle\frac{d\sigma}{d\omega}(np\to \Sigma_c^0\overline{D^0} p),$$ 
$$ \displaystyle\frac{d\sigma}{d\omega}(np\to \Sigma_c^+ D^- p)=\displaystyle\frac{d\sigma}{d\omega}(np\to \Sigma_c^+\overline{D^0} n).$$ 
The normalization implicitely taken here is such that:
$$\displaystyle\frac{d\sigma}{d\omega}=|f_{01}|^2+|f_{10}|^2+2|f_{11}|^2$$
for any process of $np$-interaction, whereas the term $|f_{01}|^2$ is missing for the cross section of any process of $pp$-interaction.
\section{Meson exchanges}

\begin{figure}
\mbox{\epsfysize=15.cm\leavevmode \epsffile{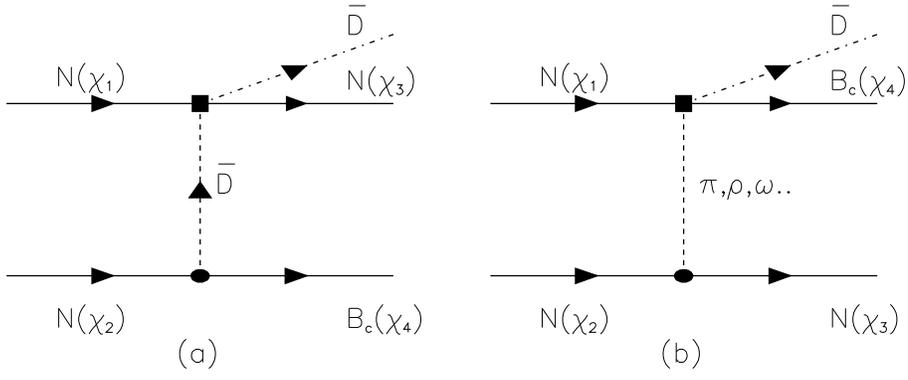}}
\vspace*{.2 truecm}
\caption{Feynman diagrams for the process 
$N+N\to {\cal B}_c+\overline{D}+N$, corresponding to two possible $t-$exchanges: (a) $\overline{D}$-meson exchange, (b) light mesons exchange.}
\label{fig:fig1}
\end{figure}

Following the ideology of meson production in $NN$-collisions, we consider different meson exchanges in $t$-channel, for different processes of associative  charm particle production $N+N\to {\cal B}_c+\overline{D}+N$ (Fig. \ref{fig:fig1}).
Taking into account the large difference in masses of the $\overline{D}$-meson and the light mesons, $\pi,\rho,\omega..$, one can expect mostly  exchanges of non-charmed mesons in $t$-channel. This is not the case for threshold production of heavy hadrons, where large values of momentum transfer squared are involved:
$$t_1=(p_2-p_4)^2=-m(M-m)-\mu M\simeq -5.5 ~\mbox{GeV}^2,~\overline{D}-\mbox{exchange},$$
$$t_2=(p_1-p_3)^2=-m(M-m)-\mu m\simeq -3.0 ~\mbox{GeV}^2,~\pi-\mbox{exchange},$$
where $M$, $m$, and $\mu$ are the masses of the charmed baryon, $N$ and  $\overline{D}$-meson, respectively. The ratio of the propagators, $(t_1-\mu^2)/(t_2-m_{\pi}^2)\simeq 3$, favors $\pi$-exchange, but the $\overline{D}$-contribution can not be neglected. This kind of kinematical estimation does not work in case of associative strange particle production, $N+N\to\Lambda(\Sigma)+K+N$, where, namely $K-$meson exchange seems more probable than $\pi$-exchange \cite{Mo02}. This is supported by, at least, two arguments: the relative value of the total cross section for $\Lambda$ and $\Sigma$-production near threshold, and the negative value of the $D_{nn}$-coefficient in the process $\vec p+p\to\vec\Lambda+K^++p$ \cite{La01}. Therefore, as starting point for the present estimation, we will consider here only $\overline{D}$-exchange.

\subsection{The reaction $p+p\to \Lambda_c^+(\Sigma_c^+)+\overline{D^0}+p$} 

\begin{figure}
\mbox{\epsfysize=15.cm\leavevmode \epsffile{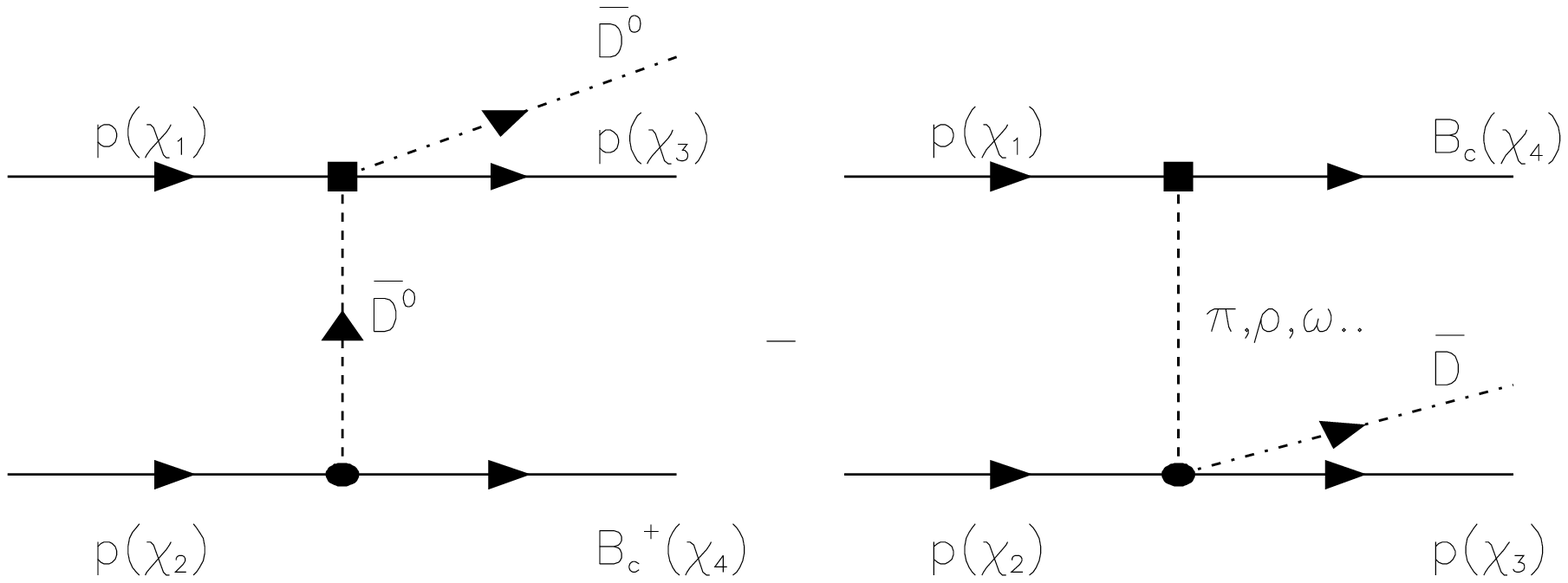}}
\vspace*{.2 truecm}
\caption{$\overline{D^0}$-exchanges for the process $p+p\to {\cal B}_c+\overline{D^0}$+p.}
\label{fig:fig2}
\end{figure}

Taking into account the identity of the colliding protons, the $D$-exchange  is described by two Feynman diagrams (Fig. \ref{fig:fig2}), with the following matrix element:
$$
{\cal M}_D= {\cal M}_{1D}-{\cal M}_{2D},$$
\begin{equation}
{\cal M}_{1D}= -\displaystyle\frac{g_{p{\cal B}_c^+\overline{D^0}}}{t_1-\mu^2}
\displaystyle\frac{k{\cal N}}{E+m}{\cal A}(\overline{D^0} p\to \overline{D^0} p)
({\chi}^{\dagger}_3 {\cal I}\chi_1)
(\chi^{\dagger}_4\vec{\sigma}\cdot\hat{\vec k } \chi_2),
\label{eq:ms1}
\end{equation}
$${\cal M}_{2D}=-{\cal M}_{1D}~(\chi_1\leftrightarrow \chi_2),$$
where ${\cal A}(\overline{D^0} p\to \overline{D^0} p)$ is the threshold  amplitude for $\overline{D^0} p$ elastic scattering, $g_{p{\cal B}_c^+ \overline{D^0}}$ is the pseudoscalar coupling constant for the vertex 
$p\to {\cal B}_c^+ +\overline{D^0}$, ${\cal N}=2M(E+m)$ is the normalization factor, due to the presence of four baryons in the considered process, the factor $k/(E+M)$ results from the transformation of the Dirac spinor product: $\overline{u}\gamma_5 u$ to its  two-component equivalent, $\chi^{\dagger}\vec\sigma\cdot\hat{\vec k } \chi$; $E$ is the threshold energy of the initial proton in the reaction CMS, ${\cal I}$ is the unit $2\times 2$-matrix.

We assume, in Eq. (\ref{eq:ms1}) the pseudoscalar nature of the $N\to {\cal B}_c^++\overline{D^0}$-vertex, i.e. the product of the parities of all these particles must be $P(N{\cal B}_c^+\overline{D^0})=-1$, following the quark model.

To transform the matrix element ${\cal M}_D$ into the "standard"  form of Eq. (\ref{eq:mnp}), we apply the Fierz transformation, in its two-component form:
\begin{eqnarray}
&(\chi^{\dagger}_3{\cal I}\chi_1)~(\chi^{\dagger}_4 \vec{\sigma}\cdot\hat{\vec k } {\chi}_2)=&
\displaystyle\frac{1}{2}\left \{
-(\tilde{\chi}_2~\sigma_y ~\chi_1)~(\chi^{\dagger}_4
\vec{\sigma}\cdot\hat{\vec k }\sigma_y\tilde{\chi}^{\dagger}_3 )+\right .\nonumber \\
&& (\tilde{\chi}_2~\sigma_y\vec{\sigma}\cdot\hat{\vec k }\chi_1)~(\chi^{\dagger}_4\sigma_y  \tilde{\chi}^{\dagger}_3 )- \label{eq:mppf} \\&&
\left . i[\tilde{\chi}_2~\sigma_y (\vec{\sigma}\times \hat{\vec k })_a\chi_1]~(\chi^{\dagger}_4 \sigma_a\sigma_y\tilde{\chi}^{\dagger}_3 )\right \}.\nonumber
\end{eqnarray}
From Eq. (\ref{eq:mppf}) one can see that ${\cal M}_{1D}$ contains not only the structures which are  allowed by symmetry selection rules, but also a contribution which corresponds to a singlet-triplet transition  (first term in Eq. (\ref{eq:mppf})). Such transition is forbidden by the generalized Pauli principle, following from the isotopic invariance of the strong interaction and should not appear in the total matrix element ${\cal M}_D$. This is an indication that only the contribution of both diagrams generates the correct answer. The expression for total matrix element ${\cal M}_D$ is:
\begin{eqnarray}
&
{\cal M}_D(pp\to {\cal B}_c^+\overline{D^0}p)= &\displaystyle\frac{g_{p{\cal B}_c^+\overline{D^0}}}{t_1-\mu^2}
\displaystyle\frac{k{\cal N}}{E+m}{\cal A}(\overline{D^0} p\to \overline{D^0} p)
\nonumber\\
&&
\left \{ - (\tilde{\chi}_2~\sigma_y ~\vec{\sigma}\cdot\hat{\vec k }\chi_1)~(\chi^{\dagger}_4\sigma_y \tilde{\chi}^{\dagger}_3 )+
 i[\tilde{\chi}_2~\sigma_y(\vec{\sigma}\times 
\hat{\vec k})_a\chi_1]~(\chi^{\dagger}_4 \sigma_a\sigma_y \tilde{\chi}^{\dagger}_3 )\right \},\label{eq:md} 
\end{eqnarray}
with the necessary threshold spin structure and with definite relation between the threshold partial amplitudes:
$$
f_{10}(pp\to {\cal B}_c^+\overline{D^0}p)=-f_{11}(pp\to {\cal B}_c^+\overline{D^0}p).$$
Comparing Eqs. (\ref{eq:mppf}) and (\ref{eq:md}), one can conclude that the interference of the two diagrams (Fig. \ref{fig:fig1}) is very important, near threshold: a correct evaluation of all terms results in a 50\% increase of the cross section, compared with calculations based on non-interfering diagrams. A coherent consideration of the two contributions is especially important for the analysis of polarization phenomena in the near threshold region.

\subsection{The reaction $n+p\to \Lambda_c^++\overline{D^0}+n$} 

\begin{figure}
\mbox{\epsfysize=15.cm\leavevmode \epsffile{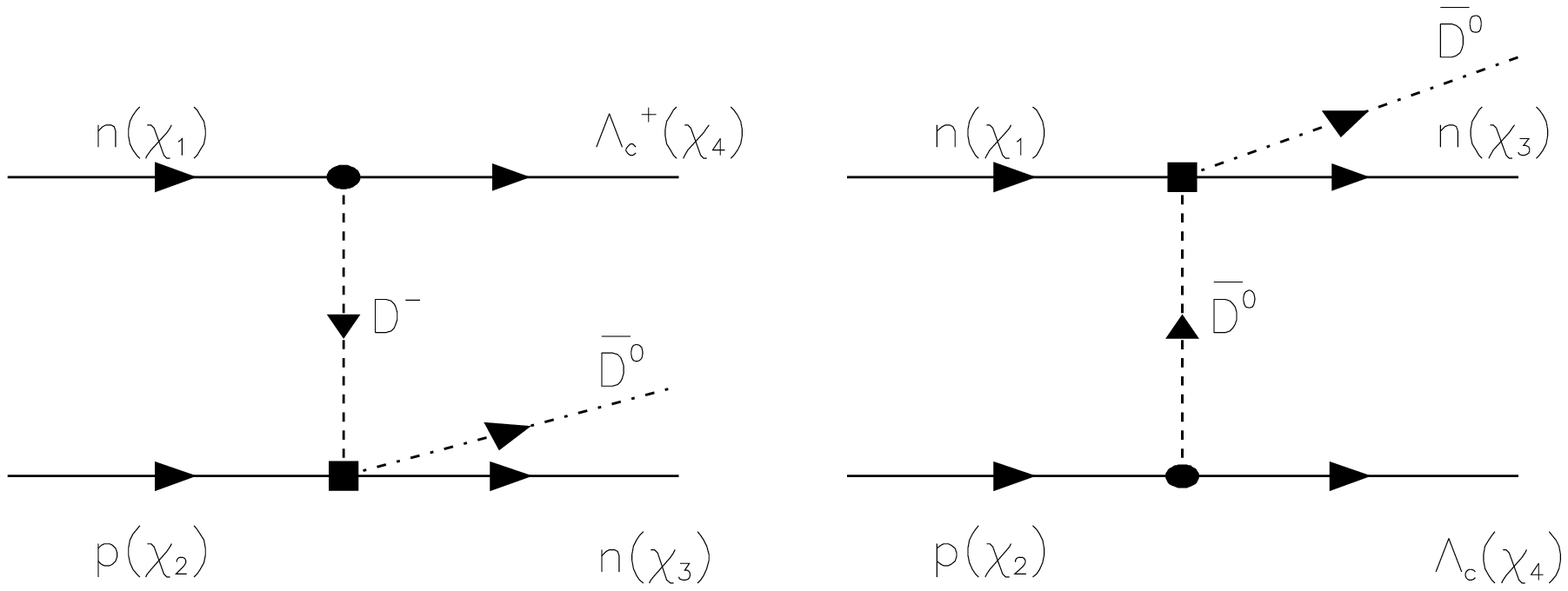}}
\vspace*{.2 truecm}
\caption{$D$-exchange for the process $n+p\to n+\Lambda_c+\overline{D^0}$.}.
\label{fig:fig3}
\end{figure}

The $\overline{D}$-exchange, here, contains two Feynman diagrams - with exchange by neutral and charged $D-$mesons (Fig. \ref{fig:fig3}).
The matrix element can be written as follows:
$$
{\cal M}_D= {\cal M}_{\overline{D^0}}-{\cal M}_{D^-}.$$
Following the two-components Feynman rules for the considered diagrams, after the corresponding Fierz transformation, one can find:
\begin{eqnarray}
{\cal M}_{\overline{D^0}}&= &-\displaystyle\frac{g_{p\Lambda_c^+\overline{D^0}}}{t_1-\mu^2}
\displaystyle\frac{k{\cal N}}{E+m}{\cal A}(\overline{D^0} n\to \overline{D^0} n)
(\chi^{\dagger}_4\vec{\sigma}\cdot\hat{\vec k }{\chi}_2)({\chi}^{\dagger}_3{\cal I}\chi_1)=\nonumber \\
&=& 
\displaystyle\frac{1}{2} \displaystyle\frac{g_{p\Lambda_c^+\overline{D^0}}}{t_1-\mu^2}
\displaystyle\frac{k{\cal N}}{E+m}{\cal A}(\overline{D^0} n\to \overline{D^0} n)
\left \{(\tilde{\chi}_2~\sigma_y ~\chi_1)~(\chi^{\dagger}_4\vec{\sigma}\cdot\hat{\vec k }\sigma_y \tilde{\chi}^{\dagger}_3 )-
\nonumber \right .\\
&& (\tilde{\chi}_2~\sigma_y\vec{\sigma}\cdot\hat{\vec k }\chi_1)~(\chi^{\dagger}_4\sigma_y  \tilde{\chi}^{\dagger}_3 )\left. +i[\tilde{\chi}_2~\sigma_y(\vec{\sigma}\times \hat{\vec k })_a\chi_1]~(\chi^{\dagger}_4 \sigma_a\sigma_y \tilde{\chi}^{\dagger}_3 )
\right \},
\end{eqnarray}
Taking into account that ${\cal M}_{D^-}$ can be derived from ${\cal M}_{\overline{D^0}}$, after the following substitutions:
$$\chi_1\leftrightarrow\chi_2,$$
$$\hat{\vec k }\leftrightarrow -\hat{\vec k },$$ 
$$ {\cal A}(\overline{D^0} n\to \overline{D^0} n)\leftrightarrow {\cal A}(D^- p\to \overline{D^0} n),$$
one can obtain the following formula, for the total matrix element:
\begin{eqnarray}
&
{\cal M}_D (np\to p\Lambda_c^+\overline{D^0})=&
\displaystyle\frac{1}{2}\displaystyle\frac{g_{p\Lambda_c^+\overline{D^0}}}{t_1-\mu^2}
\displaystyle\frac{k{\cal N}}{E+m}\nonumber \\
&& \left \{
(\tilde{\chi}_2~\sigma_y ~\chi_1)~(\chi^{\dagger}_4
\vec{\sigma}\cdot\hat{\vec k }\sigma_y \tilde{\chi}^{\dagger}_3 )
\left [{\cal A}(\overline{D^0} n\to \overline{D^0} n)-{\cal A}(D^- p\to \overline{D^0} n)\right ]+\right .
\nonumber \\
&& \left [-(\tilde{\chi}_2~\sigma_y \vec{\sigma}\cdot\hat{\vec k }~\chi_1)~(\chi^{\dagger}_4\sigma_y \tilde{\chi}^{\dagger}_3 )
+i\left (\tilde{\chi}_2~\sigma_y (\vec{\sigma}\times \hat{\vec k } )_a\chi_1\right )(\chi^{\dagger}_4 \sigma_a\sigma_y \tilde{\chi}^{\dagger}_3 )\right ]   \nonumber \\
&&\left . \left [{\cal A}(\overline{D^0} n\to \overline{D^0} n)+{\cal A}(D^- p\to \overline{D^0} n)\right ]\right \}.\label{eq:mdd}
\end{eqnarray}
Using the isotopic invariance for the processes of $\overline{D} N$-scattering,
$${\cal A}(\overline{D^0} n\to \overline{D^0} n)-{\cal A}(D^- p\to \overline{D^0} n)=a_0,$$
$${\cal A}(\overline{D^0} n\to \overline{D^0} n)+{\cal A}(D^- p\to \overline{D^0} n)=a_1,$$
where $a_0$ and $a_1$ are the $\overline{D} N$-scattering lengths, corresponding to the possible values of the total isospin for $\overline{D} N$-scattering, one can sees that the $f_{01}$-amplitude for $n+p\to \Lambda_c^++\overline{D^0}+n$ depends only on $a_0$, and both triplet amplitudes, $f_{10}$ and $f_{11}$, depend on $a_1$ - in agreement with the correlation between the spin and the isospin structures discussed above. Note that, despite the large mass of the $D$-meson, the scattering lengths $a_0$ and $ a_1$ can be considered real quantities, as in case of $KN$-scattering, for the same physical reason: following the unitarity condition there is no intermediate state for the $\overline{D} N$-channel at zero $\overline{D}$-meson energy, due to the absence of baryons with negative charm.

The expressions (\ref{eq:md}) and (\ref{eq:mdd}) allow to write the two isotopic amplitudes $A_1$ and $A_0$, for $\Lambda_c^+ \overline{D^0}$-production in $NN$-collisions:
\begin{eqnarray*}
&A_1=&a_1\displaystyle\frac{g_{p\Lambda_c^+\overline{D^0}}}{t_1-\mu^2}
\displaystyle\frac{k{\cal N}}{E+m} \\
&& \left \{ -(\tilde{\chi}_2~\sigma_y \vec{\sigma}\cdot\hat{\vec k }~\chi_1)~(\chi^{\dagger}_4\sigma_y \tilde{\chi}^{\dagger}_3 )
+i\left [\tilde{\chi}_2~\sigma_y (\vec{\sigma}\times \hat{\vec k })_a\chi_1\right ](\chi^{\dagger}_4 \sigma_a\sigma_y \tilde{\chi}^{\dagger}_3 )\right \}  \\
&A_0=&a_0\displaystyle\frac{g_{p\Lambda_c^+\overline{D^0}}}{t_1-\mu^2}
\displaystyle\frac{k{\cal N}}{E+m}
(\tilde{\chi}_2~\sigma_y \chi_1)
(\chi^{\dagger}_4\vec{\sigma}\times \hat{\vec k }\sigma_y \tilde{\chi}^{\dagger}_3 ). 
\end{eqnarray*}
These expressions allow us to find the ratio  ${\cal R}_D$ of the cross sections of $\Lambda_c\overline{D}$-production in $np-$ and $pp$-collisions:
\begin{equation}
{\cal R}_D=\displaystyle\frac{\sigma(np\to\Lambda_c^+\overline{D^0} n)}
{\sigma(pp\to\Lambda_c^+\overline{D^0} p)}=
\displaystyle\frac{\sigma(np\to\Lambda_c^+{D^-} p)}
{\sigma(pp\to\Lambda_c^+\overline{D^0} p)}=\displaystyle\frac{1}{4}
\left(1+\displaystyle\frac{a_0^2}{3a_1^2}\right )>\displaystyle\frac{1}{4},
\label{eq:ratio}
\end{equation}
where the index $D$ underlines that this prediction is valid only in framework of $D$-exchange.

This result can be modified by the presence of $\Lambda_c N$-interaction in S-state.  Taking into account the singlet $a_s(\Lambda_c N)$ and triplet $a_t(\Lambda_c N)$ scattering lengths, a simple estimation of this effect gives:
\begin{equation}
{\cal R}_D^{FSI}={\cal R}_D+\displaystyle\frac{1}{6}(1-x_{ts})\displaystyle\frac{a_0^2}{a_1^2},
\label{eq:rfsi}
\end{equation}
where $x_{ts}=a_t^2(\Lambda_c N)/a_s^2(\Lambda_c N)$, i.e. FSI will decrease (increase) the ratio ${\cal R}_D$ when the triplet $\Lambda_c N$ scattering is stronger (weaker) than the singlet $\Lambda_c N$ scattering. In any case  ${\cal R}$ is sensitive to FSI interaction. In the initial state, taking into account that the reaction threshold is quite large,  the $pp$- and $np$-interactions can be assumed similar. Therefore one expects that ISI, for the calculation of 
${\cal R}_D$, are less important than in the case of strange particle production. Note that $|a_0/a_1|\simeq 0.1$ for elastic $KN$-scattering, and it is likely that this estimation is correct also for elastic $\overline{D}N$-scattering. In this context the ratio (\ref{eq:rfsi}) will be insensitive to $\Lambda_c N$ FSI-interaction also.

\section{Estimation of the cross section}

We give here an estimation of the cross section for $N+N\to \Lambda_c +\overline{D}+N$ in the threshold region, where the energy dependence of the total cross section can be written as \cite{Mo02}:

\begin{equation}
\sigma_{pp\to\Lambda_c D N}=\overline{|{\cal M}_D|^2}V_{ps}/{\cal F}, 
\label{eq:st1}
\end{equation}
with 
\begin{equation}
V_{ps}=\displaystyle\frac{\pi^3}{2}\displaystyle\frac{(mM\mu)^{1/2}}{(m+M+\mu)^{3/2}} Q^2,~Q=\sqrt{s}-(m+M+\mu),~{\cal F}=4\sqrt{s} k(2\pi)^5,
\label{eq:st2}
\end{equation}
where we assumed, for simplicity, that $\overline{|{\cal M}|_D^2}$ is constant, in the region of applicability of $Ss$-production of final particles.

Let us consider the ratio of the threshold matrix elements for the processes 
$p+p\to\Lambda_c^+ +\overline{D^0} +p$ ($D-$exchange) and $p+p\to\Lambda +K^+ +p$ ($K-$exchange). One can find:
\begin{equation}
{\cal R}_{DK}=\displaystyle\frac{\overline{|{\cal M}_D|^2}}{\overline{|{\cal M}_K|^2}}=
\left (\displaystyle\frac{g_{p\Lambda_c^+\overline{D^0}}}
{g_{p\Lambda K^+}} \right )^2\left [\displaystyle\frac{a_1(DN)}{a_1(KN)}\right ]^2
\left (
\displaystyle\frac{t_K-m_K^2}{t_D-m_D^2}\right )^2
\displaystyle\frac{ M(\Lambda_c)}{ M(\Lambda)}
\displaystyle\frac{(E-m)_D}{(E-m)_K}, 
\label{eq:eq12}
\end{equation}
where $(E-m)_{D,K}=\displaystyle\frac{1}{2}(M-m+\mu)$, $M(\mu)$ is the mass of $\Lambda_c (D)$ for $D$-production and $M(\mu)$ is the mass of $\Lambda (K)$ for $K$-meson production. Taking the numerical values: 
\begin{equation}
M(\Lambda_c)/M(\Lambda)\simeq 2.05,~{(E-m)_D}/{(E-m)_K}=4.79, ~({t_K-m_K^2})^2/({t_D-m_D^2})^2\simeq 1.1\cdot 10^{-2}
\label{eq:aqq}
\end{equation}
one can find:
\begin{equation}
{\cal R}_{DK}=0.11 \left (\displaystyle\frac{g_{p\Lambda_c^+\overline{D^0}}}
{g_{p\Lambda K^+}}\right )^2 \left [\displaystyle\frac{a_1(DN)}{a_1(KN)}\right ]^2.
\label{eq:eq13}
\end{equation}
Therefore, the ratio of total cross sections, at a fixed value of $Q$, for 
$\Lambda_c^+\overline{D^0}$ and $\Lambda K^+$ production in $pp$-collisions can be written as:
\begin{eqnarray}
&\displaystyle\frac{\sigma_t(pp\to\Lambda_c^+ \overline{D^0}p)}{\sigma_t(pp\to\Lambda K^+p)}=&{\cal R}_{DK}\left [\displaystyle\frac{M(\Lambda_c) \mu(\overline{D^0})}{M(\Lambda)\mu(K)}\right ]^{1/2}\left  [
\displaystyle\frac{m+M(\Lambda)+ \mu(K)}{m+M(\Lambda_c)+ \mu(\overline{D^0})}\right ]^{5/2}
\displaystyle\frac{k(\Lambda)}{k(\Lambda_c)}\nonumber \\
&&
=0.27 {\cal R}_{DK}=2.98\cdot10^{-2} \left (\displaystyle\frac{g_{p\Lambda_c^+\overline{D^0}}}
{g_{p\Lambda K^+}}\right )^2\left [\displaystyle\frac{a_1(DN)}{a_1(KN)}\right ]^2.
\label{eq:nr}
\end{eqnarray}

We neglected, in this estimation, an important ingredient, the phenomenological hadron form factors, which are traditionally included in these calculations. Such form factors will suppress the charmed particle production near threshold, according to:
\begin{equation}
{\cal D}_{F}= \left (1-\displaystyle\frac{t_K-m_K^2}{\lambda^2}\right) ^2/
\left (1-\displaystyle\frac{t_D-m^2_D}{\lambda^2}\right) ^2,
\label{eq:ff}
\end{equation}
where, for simplicity, each hadronic vertex is described by the same monopole form factor, with the same cut-off parameter $\lambda$, $t_K$ and $t_D$ are the corresponding  momentum transfer squared in $K$- or $D$-meson threshold production. Taking, as an example, $\lambda$=1.5 GeV, one can find a suppression factor ${\cal D}_{F}\simeq 0.08$.

In framework of $SU_4$-symmetry, the coupling constants ${g_{p\Lambda_c^+\overline{D^0}}}$ and ${g_{p\Lambda K^+}}$ are equal, so finally:
\begin{equation}
\displaystyle\frac{\sigma_t(pp\to\Lambda_c^+ \overline{D^0}p)}{\sigma_t(pp\to\Lambda K^+p)} \simeq 3\cdot 10^{-2} \left [\displaystyle\frac{a_1(DN)}{a_1(KN)}\right ]^2.
\label{eq:eq14}
\end{equation}
Dipole hadronic form factors will result in a cross section one order of magnitude 
smaller, for $\Lambda_c^+ \overline{D^0}$-production. 
On the basis of Eq. (\ref{eq:eq14}), taking the experimental data on the total cross section for $p+p\to \Lambda+ K^++p$ \cite{Mo02}, one can predict the following energy dependence for the cross section of open charm production:
\begin{equation}
\sigma(pp\to\Lambda_c^+ \overline{D^0}p)\simeq 0.2(Q/0.1 ~\mbox{GeV})^2\mbox{ nb}.
\label{eq:eq15}
\end{equation}
As we mentioned above, no experimental data exist in the threshold region. The lowest energy where the open charm cross section has been measured, $E_p$=70 GeV, $\sqrt{s}=11.46$, corresponds to $Q=6$ GeV. From Eq.
(\ref{eq:eq15}) we find, for this energy: $\sigma(pp\to\Lambda_c^+ \overline{D^0}p)\simeq 0.7$ $\mu$b, to be compared to the experimental value $\sigma(pp\to~charm )=1.6^{+1.1}_{-0.7}~ (stat)\pm 0.3~ (syst)$ $\mu$b \cite{Am01}. However, this apparent agreement, resulting form a very large extrapolation in $Q$, can be accidental, because Eqs. (\ref{eq:st1}) and (\ref{eq:st2}) rigorously hold only for $Q< \mu$.
\section{Conclusions}
In this paper we analyzed the simplest processes of open charm production in the near threshold region of $NN$-collisions, $N+N\to \Lambda_c(\Sigma_c)+\overline{D}+N$, in a model independent way, using the symmetry properties of the strong interaction, in analogy with strange particle production \cite{Re97}.  We established the spin and isospin structure of the corresponding threshold matrix element in exact form and showed a strong correlation of spin and isospin variables.

The $D$-exchange mechanism for $N+N\to \Lambda_c(\Sigma_c)+\overline{D}+N$ can be considered a natural generalization of the $K$-exchange for strange particle production, which appears as the main mechanism in the near threshold region.

We have described the open charm production in terms of the following ingredients:
\begin{itemize}
\item the coupling constants  $g_{N\Lambda_c\overline{D}}$ and $g_{N\Sigma_c\overline{D}}$ and the threshold amplitude for the elementary
$\overline{D} N$-scattering (both these quantities can be related to the corresponding values for strange particles, namely the $g_{N\Lambda K}$ coupling constant and the $KN$-scattering length, through $SU_4$-symmetry);
\item the parameters (scattering lengths and effective radii) of final strong $N\Lambda_c$-interaction described in analogy with low energy $\Lambda N$-interaction;
\item the hadronic phenomenological form factors.
\end{itemize}
This simple and transparent model allows us to predict the relative value of the cross sections for $\Lambda_c\overline{D}$- and $\Lambda K$-production, at the same value of the energy excess $Q$. We find a difference of three orders of magnitude in the values of the corresponding cross sections, on the basis of simple kinematical arguments, such as the difference of the reaction thresholds and of the meson propagators (due to the difference of the threshold momentum transfer). Phenomenological hadron form factors also play a large role in the threshold region of charm production.

\end{document}